\documentclass[twoside]{LCWS11}
\usepackage[latin1]{inputenc}
\usepackage[dvips]{graphicx,epsfig,color}
\usepackage{wrapfig,rotating}
\usepackage{amssymb,amsmath,array}
\usepackage{cite}
\def\bmag#1{{|{\mathbf #1}|}}

\def\Dsl{\hbox{/\kern-.6000em D}} %roman D

\def\dsl{\,\raise.15ex\hbox{/}\mkern-13.5mu D}

\def\ltap{\ \raise.3ex\hbox{$<$\kern-.75em\lower1ex\hbox{$\sim$}}\ }
\def\gtap{\ \raise.3ex\hbox{$>$\kern-.75em\lower1ex\hbox{$\sim$}}\ }
\def\OMIT#1{}

\def\lsim{\mathrel{\raise.3ex\hbox{$<$\kern-.75em\lower1ex\hbox{$\sim$}}}}
\def\gsim{\mathrel{\raise.3ex\hbox{$>$\kern-.75em\lower1ex\hbox{$\sim$}}}}

\newcommand{\bmp}{\mathbf p}

\newcommand{\braket}[1]{\langle #1 \rangle}
\newcommand{\Tprod}[1]{{\mathrm T}\lbrack #1 \rbrack}
\newcommand{\grtsim}{\mbox{\raisebox{-3pt}{$\stackrel{>}{\sim}$}}}

\begin{document}
\title{
%%%%   Paper title goes here  %%%%%%%%%%%%%%
Non-resonant effects in the $t\bar{t}$ resonance region} %% 
%***********************************************************************
% AUTHORS INFORMATION AREA
%***********************************************************************
\author{Pedro Ruiz-Femen\'\i a$^{1,2}$
% Optional short acknowledgment: remove next line if non-needed
\thanks{Talk based on work done in collaboration with A. Hoang and C. Reisser, and on
work done in collaboration with M. Beneke and B. Jantzen. Preprint numbers: UWThPh-2012-10, IFIC/12-12.  }
% DO NOT MODIFY THE FOLLOWING '\vspace' ARGUMENT
\vspace{.3cm}\\
% Addresses and institutions (remove "1- " in case of a single institution)
1- University of Vienna - Faculty of Physics \\
Boltzmanngasse 5, A-1090 Wien, Austria 
%% Remove the next three lines in case of a single institution
\vspace{.1cm}\\
2- Instituto de F\'\i sica Corpuscular (IFIC), CSIC-Universitat de Val\`encia \\
Apdo. Correos 22085, E-46071 Valencia  - Spain\\
}
%%***********************************************************************
% END OF AUTHORS INFORMATION AREA
%***********************************************************************

\maketitle

\begin{abstract}
The recent developments on the computation of non-resonant corrections to the $e^+e^-\to W^+W^-b\bar{b}$
cross section in the top-antitop resonance region are reviewed in this talk. Non-resonant production of the
final state $W^+W^- b\bar{b}$ 
starts to contribute at NLO in the nonrelativistic power-counting $v ~\sim \alpha_s \sim \sqrt{\alpha_{\rm EW}}$.
The corrections induced by non-resonant effects reduce the cross section in the top-antitop resonance peak region at a level
which is comparable to the expected experimental precision at the future linear collider, and are thus relevant
for a high-precision top mass and width determination. 
\end{abstract}

\section{Introduction}

A next generation $e^+e^-$ linear collider such as the International Linear Collider (ILC) operating
at energies around the top-antitop threshold will allow us to measure the top-quark's 
properties with unprecedented precision.
The top-quark mass 
is currently known from direct production at the 
Fermilab Tevatron (and soon at the Large Hadron Collider) with 
a precision $\grtsim 1$~GeV. From a 
threshold scan of the $e^+ e^-\to t\bar t$ cross section 
at the ILC, however, 
an order of magnitude improvement in the precision can be achieved 
experimentally~\cite{Martinez:2002st}. Aside from determining a 
fundamental parameter of the Standard Model, its precise knowledge is
important for precision tests of the Standard Model and its extensions. 
Other characteristics of the top quark 
such as its width and Yukawa coupling provide information 
about its coupling to other particles and the mechanism of 
electroweak symmetry breaking. 
For these reasons top-quark pair 
production near threshold in $e^+ e^-$ annihilation has been 
thoroughly investigated following the 
nonrelativistic QCD (NRQCD) approach,
which treats the leading colour-Coulomb force between
the nearly on-shell top and antitop exactly to all orders 
in perturbation theory.  
In this framework, where the strong 
coupling $\alpha_s$ is of the same order as $v$,
the small relative velocity of the top and antitop, 
QCD corrections to the total cross section are known
at the next-to-next-to-next-to-leading order (NNNLO)~\cite{Beneke:2005hg,Beneke:2008ec,Beneke:2008cr}, and 
higher-order logarithms $(\alpha_s \log v)^k$ 
have been resumed to next-to-next-to-leading logarithmic order (NNLL)~\cite{Hoang:2000ib}.
The NNLL prediction, which had a normalization uncertainty in the
total cross section $\gsim 6\%$, has been recently updated~\cite{Hoang:2011it}, showing
a substantial reduction of the renormalization scale dependence. 
An analysis collecting all the pieces that contribute to the 
full NNNLO total cross section is still pending (see~\cite{Beneke:2008ec} for the latest
summary), which would provide information about the convergence of the
non-logarithmic piece of the QCD corrections.

Here we focus on subleading corrections of electroweak origin.
The top quark is unstable with a 
significant width $\Gamma_t$ of about $1.5\,$GeV due to the electroweak 
interaction.
Once the top width is included, due to top decay, the physical 
final state is $W^+ W^- b\bar b$ -- at least 
if we neglect the decay of top into strange and down quarks, as 
justified by $V_{tb}\approx 1$, and consider $W$ bosons as 
stable. The $W^+W^- b \bar{b}$ final state can also be produced 
non-resonantly, {\it i.e.} through processes which do not involve  a nearly on-shell
$t\bar{t}$ pair.   
The latter effects are not included in the standard
nonrelativistic treatment used to compute the dominant QCD corrections,
which can only describe the resonant process
with a nearly on-shell $t\bar{t}$ intermediate state, but can be accommodated
using the effective field theory (EFT)
formalism to describe pair production of unstable particles at threshold~\cite{Beneke:2003xh,Hoang:2004tg,Beneke:2007zg}. 
Adopting a counting scheme where 
$\alpha_{\rm EW}\sim \alpha_s^2$, the leading 
non-resonant and off-shell effects are NLO for 
the total cross section, since there is an additional power 
of $\alpha_{\rm EW}$ but no phase space suppression, hence 
the relative correction is $\alpha_{\rm EW}/v\sim \alpha_s$.
At NLO they can be classified as part of the electroweak corrections
to the $e^+e^-\to W^+W^- b\bar{b}$
cross section, and are dominant compared to the purely resonant electroweak effects, 
which first contribute at NNLO~\cite{Hoang:2004tg}. The identification
of non-resonant production of the final state $W^+W^-b\bar{b}$ as
a pure electroweak effect does not hold anymore at NNLO, where gluon corrections
to the non-resonant part have to be considered. 

In this talk I discuss the status of the calculation of the non-resonant 
contributions to the $e^+ e^-\to W^+ W^- b\bar b$ process 
in the $t\bar t$ resonance 
region. The results at NLO are fully known for the total cross section as well as including 
invariant-mass cuts on the $Wb$ pairs. 
At NNLO, interesting conceptual issues regarding the interplay between resonant and non-resonant
parts of the calculation arise. These are discussed in  Sec.~\ref{sec:NNLO}.
The computation of the full set of NNLO non-resonant corrections represents
a much more difficult task that has not yet been attempted. 
An alternative approach, named ``NRQCD phase space matching'',
to compute the non-resonant effects beyond NLO 
entirely through calculations in NRQCD has been proposed~\cite{Hoang:2008ud,Hoang:2010gu},
which works if moderate invariant-mass cuts on
the $Wb$ pairs are applied. The concepts of the phase space matching and the
results from this method are reviewed in the last section.

\section{Non-resonant NLO electroweak contributions}
\label{sec:NLO}

The cross section for the 
$e^+ e^-\to W^+ W^- b\bar b$ process  is obtained from the
$W^+bW^-\bar{b}$ cuts of the $e^+e^-$ forward-scattering amplitude. 
In the energy region $\sqrt{s}\approx 2 m_t$ 
 the amplitude is 
dominated by the production of resonant top quarks with small 
virtuality. A separation of resonant and non-resonant effects
can be consistently achieved within the unstable-particle effective field 
theory~\cite{Beneke:2003xh}. This allows us to integrate out hard modes ($\sim m_t$)
 and represent the forward-scattering amplitude as the 
sum of two terms, 
\begin{eqnarray}
\label{eq:master}
i {\cal A} &=&\sum_{k,l} C^{(k)}_p  C^{(l)}_p \int d^4 x \,
\braket{e^- e^+ |
\Tprod{i {\cal O}_p^{(k)\dagger}(0)\,i{\cal O}_p^{(l)}(x)}|e^- e^+} 
\nonumber\\
&&+ \,\sum_{k} \,C_{4 e}^{(k)} 
\braket{e^- e^+|i {\cal O}_{4e}^{(k)}(0)|e^- e^+}. \;\;\;\;\;\;
\vspace*{-0.cm}
\end{eqnarray}
\begin{figure}[t]
%\begin{center}
\hskip -.5cm
\includegraphics[width=1.1\textwidth]{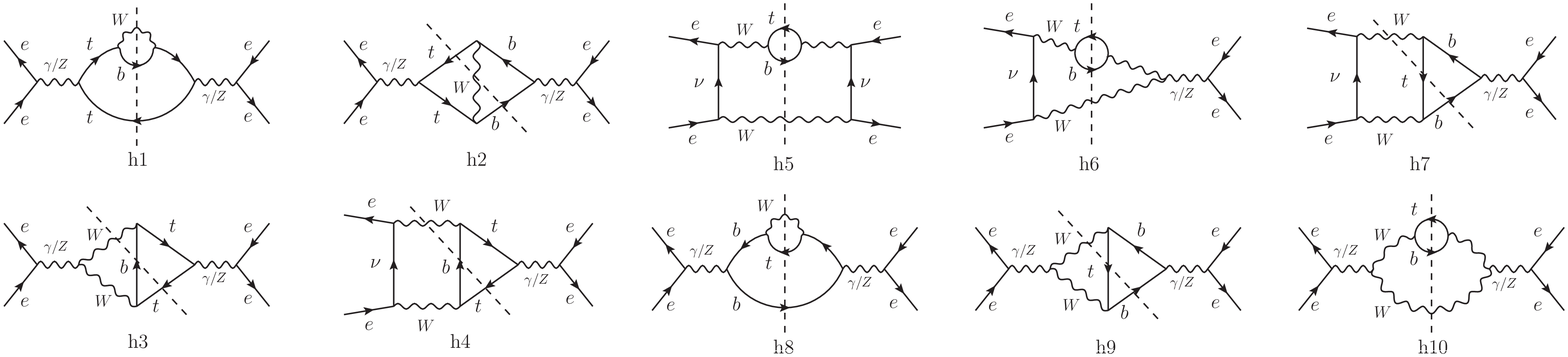}
% \vspace*{0.2cm}
\caption{Two-loop forward-scattering amplitude diagrams with 
$\bar{t} b W^+$ cuts. $t \bar{b} W^-$ cuts and symmetric diagrams 
are not shown.}
\label{fig1}
%\end{center}
\end{figure}

The matrix elements in (\ref{eq:master}) are evaluated in the 
``low-energy'' effective theory, which includes elements of 
soft-collinear and nonrelativistic effective theory. The first term 
on the right-hand side of (\ref{eq:master}) describes the production 
of a resonant $t\bar t$ pair in terms of production (decay)
operators ${\cal O}_p^{(l)}(x)$ (${\cal O}_p^{(k)\dagger}(x)$) 
with short-distance coefficients $C^{(k,l)}_p$. The
second term accounts for the remaining non-resonant contributions,
which in the effective theory are described by four-electron production-decay
operators ${\cal O}^{(k)}_{4e}$. 
The coefficients $C^{(k)}_{4e}$ originate from the hard contributions of
the $e^+e^-$ forward-scattering amplitude. The hard momentum region expansion dictates that the 
top-quark self-energy insertions are treated perturbatively, 
since the top lines are formally far off-shell, 
$p_t^2-m_t^2 \sim {\cal O}(m_t^2) \gg \Sigma(p_t^2)$. Accordingly the 
calculation of the coefficients $C_{4e}^{(k)}$ is performed in 
fixed-order perturbation theory in the full electroweak theory 
with no resummation of self-energy insertions in the top-quark 
propagator~\cite{Beneke:2007zg} and supplemented with an expansion 
of the amplitudes near threshold (in $\delta=s/(4m_t^2)-1 \sim v^2$).
The leading 
imaginary parts of $C^{(k)}_{4e}$ arise from the cut two-loop
diagrams of order $\alpha_{\rm EW}^3$ shown in Fig.~\ref{fig1}. The corresponding contribution to the cross section is
\vskip -0.1cm
\begin{equation}
\label{eq:nonres}
\sigma_{\rm{non-res}} = \frac{1}{s}\,
\sum_k \,\mbox{Im}\left[C_{4 e}^{(k)}\right]\,
\braket{e^- e^+|i {\cal O}_{4e}^{(k)}(0)|e^- e^+}.
\end{equation}
Technically, this simply amounts to the calculation of the 
spin-averaged tree-level processes $e^+ e^- \to t W^- \bar b$ and 
$e^+ e^- \to \bar t W^+ b$ with no width supplied to the 
intermediate top-quark propagators. Instead, the divergence 
from the top-quark propagators going on-shell is regularized
dimensionally. Details on the computation and integral representations
of the result for (\ref{eq:nonres}) can be found in~\cite{Beneke:2010mp}.
The results of~\cite{Beneke:2010mp} for the NLO non-resonant contribution to the total
cross section have been confirmed recently 
by an independent calculation~\cite{Penin:2011gg}. The latter
is based on an expansion in the parameter $\rho=1-M_W/m_t\approx0.53$,
which for individual diagrams requires to consider several orders in $\rho$ or to use
Pad\'e approximants to reach a precise numerical agreement with the integral representation of~\cite{Beneke:2010mp}.
For the sum of all diagrams, however, the leading order in $\rho$ gives an
approximation which differs from the exact result by less than 5\%. 

Through the computation of the four-electron matching coefficients loose cuts ($\sim m_t$) on the
$bW^+$ and $\bar{b}W^-$ invariant masses can be incorporated easily, as it has
been discussed in the context of $W$-pair production near threshold~\cite{Actis:2008rb}.
The result  obtained in~\cite{Beneke:2010mp} covers the case of symmetric cuts on the invariant mass of the $bW$
subsystems  ($p^2_{bW}$)  of the form
\begin{align}
\label{eq:invarmasscuts}
 m_t -\Delta M_t   \le \sqrt{p^2_{bW}} \le m_t +\Delta M_t \,,
\end{align}
% $m_t -\Delta M_t   \le \sqrt{p^2_{bW}} \le m_t +\Delta M_t$,
for $\Delta M_t\gg \Gamma_t$, up to the total cross section ($\Delta M_{t,\rm max}=m_t-M_W$).

%%%%%%%%%%%%%%%
\subsection{Results}
%%%%%%%%%%%%%%%

%  

\begin{figure}[t]
  \begin{center}
  \includegraphics[width=0.6\textwidth]{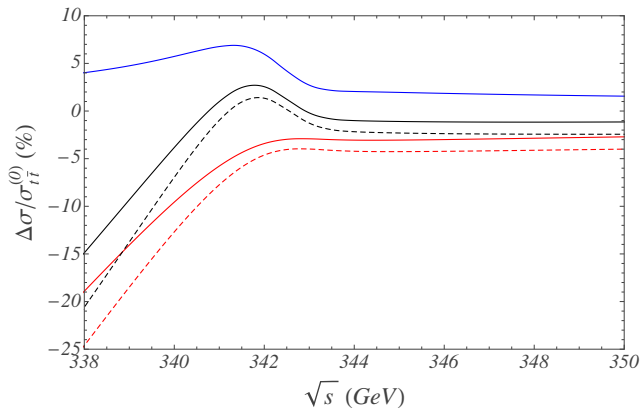}
%  \vspace*{0.2cm}
  \caption{Relative sizes of the QED and non-resonant corrections with respect the $t\bar{t}$ LO cross 
section in percent: $\sigma^{(1)}_{\rm QED}/\sigma^{(0)}_{t\bar t}$ (upper solid blue line),
$\sigma^{(1)}_{\text{non-res}}/\sigma^{(0)}_{t\bar t}$ for the total cross section
(lower solid red line)
and $\Delta M_t=15$~GeV (lower dashed red line). The relative size
of the sum of the QED and non-resonant corrections is represented by the middle lines,
for $\Delta M_{t,\rm max}$ (solid) and $\Delta M_t=15$~GeV (dashed). 
We have chosen $\alpha_s(30\,\rm{GeV})=0.142$ and $m_t=172$~GeV.
}
  \label{fig2}
  \end{center}
\end{figure}

The plot in Fig.~\ref{fig2} displays the relative sizes of the NLO electroweak
corrections with respect 
to the LO result 
for the $e^+e^-\to W^+W^-b\bar{b}$ cross section, which includes 
the summation of Coulomb corrections. 
The QED contribution represents a correction of about 2\% above threshold 
and rises to a maximum of 7\% just below the peak, while
the non-resonant contributions
give a constant negative shift of about 3\% above threshold. Below 
threshold the relative size of the non-resonant corrections is very large, 
since the LO result rapidly vanishes, reaching up to 19\%. Hence below 
threshold they represent the leading electroweak correction
to the total $t\bar{t}$ cross section. 
We observe a partial cancellation of the QED and
non-resonant corrections in the peak region and at energies above.
A sensitivity to the invariant-mass cut $\Delta M_t$ in the $bW^+$ and 
$\bar{b}W^-$ subsystem enters first at NLO through the non-resonant 
contributions. Restricting the available phase space for the final-state 
particles by tightening the invariant-mass cuts $\Delta M_t$
makes the non-resonant contributions even more important. This is shown by 
the dashed lines in Fig.~\ref{fig2}, corresponding to 
$\Delta M_t=15$~GeV. The non-resonant correction amounts
to a negative shift of 27--35~fb for $\sqrt{s}$ in the interval $(338,350)$~GeV.

\section{Non-resonant effects beyond NLO}
\label{sec:NNLO}

\subsection{Finite-width divergences}

There is an interesting conceptual issue concerning the resonant part of the 
QCD calculation of the $t\bar t$ cross section. At NNLO it 
exhibits an uncanceled ultraviolet divergence (here regulated 
dimensionally) 
\begin{equation}
\sigma_{t \bar t} \propto \frac{\alpha_s\Gamma_t}{\epsilon} 
\propto \frac{\alpha_s \alpha_{\rm EW}}{\epsilon}
\,,
\label{eq:resdiv}
\end{equation}  
which arises from the logarithmic overall divergence in the 
two-loop nonrelativistic correlation function, whose imaginary part 
gives the cross section. The overall divergence is polynomial 
in the nonrelativistic energy $E$ of the top quarks, but 
contributes to the cross section if $\Gamma_t\ne 0$, since the correlation 
function is evaluated at complex values $E\to E+i\Gamma_t$.  
%Likewise, the absorptive parts in the NRQCD matching conditions  
%coming from $Wb$ cuts in the full theory at NNLO 
%(like the example shown in Fig.~\ref{fig22})
%yield similar UV divergences in the forward scattering amplitude, as
%noted in Ref.~\cite{Hoang:2004tg}, which would not exist either for
%a stable top quark. 
It can be shown that these UV-divergences originate because the unstable 
particle propagators describing the top and antitop in the EFT allow for contributions
to the forward scattering amplitude from 
intermediate states which have arbitrarily large invariant masses. 

\begin{figure}[t]
\begin{center}  
\includegraphics[width=0.6\textwidth]{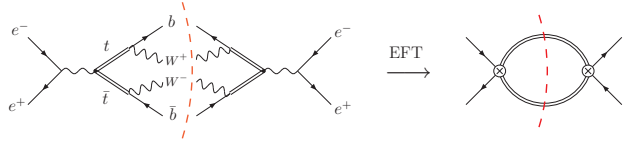}
\caption{Tree-level double-resonant diagrams in the full and effective 
  theories. Double lines are used for representing top quarks. The (red) dashed
  line denotes that we extract the imaginary part of the forward scattering amplitude
  or, equivalently, that we perform the phase space integration over the
  particles in the cut.} 
  \label{fig3}
  \end{center}
\end{figure}

The inclusion of a finite width in the top propagator changes the high-energy behaviour
of its imaginary part and makes the phase space integration extend to infinity. 
To illustrate the latter consider the Born cross section diagram in the EFT, Fig.~\ref{fig3}. 
The cutting rules for the top lines imply extracting the real part of the NRQCD propagator:
in the stable case ($i\Gamma_t/2\to i\epsilon$) we have 
\begin{equation}
\mathrm{Re} \left[ \frac{i}{p_0-\frac{\bmp^2}{2m_t}+i\epsilon} \right] = \pi\,\delta\Big(p^0-\frac{\bmp^2}{2m_t}\Big) \,,
\end{equation}
which imposes a fixed dispersion relation $p^0=\bmp^2/2m_t$,
while the real part of the unstable propagator,
\begin{equation}
\mathrm{Re} \left[ \frac{i}{p_0-\frac{\bmp^2}{2m_t}+i\frac{\Gamma_t}{2}} \right] =
\frac{\Gamma_t/2}{(p_0-\frac{\bmp^2}{2m_t})^2+ (\frac{\Gamma_t}{2})^2} \,,
\label{unstableprop}
\end{equation}
yields a Breit-Wigner distribution with support in the entire $(p^0,\bmag p)$ plane. 

From the point of view of the full theory, already taking the nonrelativistic limit makes the 
phase space integration extend to infinity. Consider the full-theory double-resonant diagram of Fig.~\ref{fig3} 
with center-of-mass momentum $q=(2m_t+E,\mathbf{0})$.
When the quark lines are close to their mass shell the 4-particle phase space integration reduces to 
\begin{equation}
\sigma_{t\bar{t}} \sim \int_{-\infty}^{+\infty} \!\!\!\!\!\! dp_0 
\int_{0}^{+\infty} \!\!\!\!\!\! d|\bmp| \bmp^2 
\frac{\Gamma_t^2}
{\left| m_t E + 2m_t p_0 -\bmp^2 + i \epsilon \right|^2
 \left| m_t E - 2m_t p_0 -\bmp^2 + i \epsilon\right|^2}
\,,
\label{eq:tree}
\end{equation}
where in the denominator we have retained the leading order term in the nonrelativistic expansion of the top and antitop off-shellness,
$(p_{t,\bar{t}}^2-m_t^2)$, with $p_{t,\bar{t}}^2=(q/2 \pm p)^2$. The $\Gamma_t^2$
factor in the numerator arises from the phase space integration of the $b W^+$ and $\bar{b}W^-$ subsystems because in the nonrelativistic limit the top and 
antitop quark are effectively on-shell.  
We notice that the integration limits in Eq.~(\ref{eq:tree}), which in the full theory 
computation are cut off by the top mass, become infinite in the nonrelativistic limit. The
boundaries of the phase space are determined by step functions in the
phase space measure, and the arguments of these functions also have to be expanded according to the nonrelativistic power-counting. 
In the limit $m_t\gg (p^0,\bmag p)$ the step functions do not depend 
on $(p^0,\bmag p)$ any longer, and are satisfied trivially, thus allowing for an infinite integration region
in these variables.
The corresponding NRQCD amplitude (diagram on the right in Fig.~\ref{fig3})  reproduces Eq.~(\ref{eq:tree})
when $\Gamma_t$ is treated as an insertion. The NRQCD power-counting, however, tells that the $\Gamma_t \sim E$
term needs to be resummed as part of the top propagator, 
thus effectively replacing $i\epsilon$ by a term proportional to $i\Gamma_t$ as in Eq.~(\ref{unstableprop}). 

Despite the integration limits, the tree-level integration in Eq.~(\ref{eq:tree}) is finite.
However the integrand becomes more sensitive to large momentum regions once
we include relativistic corrections $\sim \bmp^2/m_t^2$. Using dimensional regularization
these subleading contributions can lead to $1/\epsilon$ singularities
if the high energy behaviour
of the EFT phase space integration is logarithmic.
The lesson from 
this is that the pure resonant result alone that is usually shown in 
the literature is inconsistent
theoretically and must be supplemented with additional short-distance information
from the systematic 
calculation of the $e^+ e^-\to W^+ W^- b\bar b$ process.

\subsection{Towards an evaluation of the NNLO non-resonant contributions}
\label{NNLOnonres}

\begin{wrapfigure}{r}{0.4\columnwidth}
%\begin{figure}
\centerline{\includegraphics[width=0.35\columnwidth]{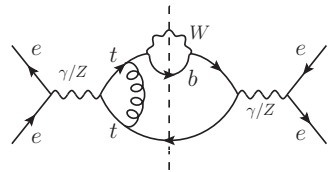}}
\caption{NNLO contribution to $\sigma_{\rm{non-res}}$}\label{fig:h1cor}
\end{wrapfigure}

Since the full-theory calculation is finite, the UV divergence in the NNLO resonant part must cancel 
with an infrared divergence that appears in the non-resonant 
term in unstable-particle effective theory from diagrams 
corresponding to off-shell top-quark decay, as discussed in \cite{Beneke:2008cr}. 
The NNLO contributions to $\sigma_{\rm{non-res}}$ are given by the
${\cal O}(\alpha_s)$ corrections to the diagrams
$h_{1-10}$ in Fig.~\ref{fig1} (an example is provided by Fig.~\ref{fig:h1cor}).
Parts of the latter consist of 
gluon radiation diagrams, which thus contribute first at NNLO in the non-resonant part,
in contrast to what happens in the resonant part, where ultrasoft gluon
radiation is a N${}^3$LO effect. 
%\end{figure}

Divergences in the non-resonant part arise when the top (or antitop) propagators go 
on-shell~\cite{Beneke:2010mp}. This is a consequence of the hard momentum
region expansion, which sets $\Gamma_t=0$ in the top-antitop propagators 
and forces to retain the leading order
term of the upper kinematic limit in 
$p_t^2\equiv p_{bW}^2$: $p_{t,\rm{max}}^2 =  m_t^2+{\cal O}(\delta)$. 
Integrating over all other kinematic variables but $p_t^2$, the non-resonant contributions
involve integrals of the form 
\begin{equation}
\int^{m_t^2}_{p_{t,\rm{min}}^2} \frac{dp_t^2}{(m_t^2-p_t^2)^{n+ a \epsilon}} \,
 = \, \frac{1}{1-n-a\epsilon}\, (m_t^2-p_{t,\rm{min}}^2)^{1-n-a\epsilon} \,,
\label{eq:ptint}
\end{equation}
where the endpoint singularity at $p_t^2=m_t^2$ is regularized in $d=4-2\epsilon$ dimensions. At NLO, only the 
diagram $h_1$ has an endpoint divergence, with $n=3/2$, and the result is therefore finite  in the limit $\epsilon \to 0$.
Gluon corrections, on the other hand, yield additional half-integer powers of $(m_t^2-p_t^2)$
and, in particular, 
$n=1$ contributions which will generate $1/\epsilon$ terms of the form (\ref{eq:resdiv}). 
For example, the gluon vertex correction
to diagram $h_1$ shown in Fig.~\ref{fig:h1cor} yields non-zero contributions with $n=2,{3 \over 2},1,{1 \over 2},\dots$, and
one can check that the resulting divergences cancel those in the resonant part of the full-theory diagram.
Clearly, the complexity 
associated with the evaluation of the full NNLO set of non-resonant corrections calls for a 
numerical implementation using Monte Carlo methods for the phase space integration.   
Such an implementation requires that the end-point divergences occurring in the 
diagrams are identified and subtracted from the amplitude, together with
the soft-collinear divergences that show up at NNLO due to gluon radiation. 
The complete set of NNLO non-resonant singularities shall be available soon~\cite{wip}.

It should be mentioned that the leading term of the expansion in $\rho=1-M_W/m_t$ of the NNLO non-resonant contribution
to the total cross section has been given in Ref.~\cite{Penin:2011gg}. The
accuracy of such an approximation, though, cannot be estimated reliably without further knowledge of the size
of subleading terms in the $\rho$-expansion.

\section{Phase space matching}
\label{sec:psmatching}

An alternative approach to account for non-resonant effects has been developed in parallel~\cite{Hoang:2008ud,Hoang:2010gu} 
that includes
the effects of invariant-mass cuts on the $Wb$ pairs entirely through calculations in an extended version of NRQCD which 
accounts for unstable particle effects systematically~\cite{Hoang:2004tg}.
In this approach, the UV-divergences from the resonant matrix elements are
compensated by imaginary counter-terms associated
with the $(e^+e^-)(e^+e^-)$ forward-scattering operators, which thus
acquire an imaginary anomalous dimension and sum large logarithms of the top
velocity. In Ref.~\cite{Hoang:2010gu} we demonstrate that for moderate top invariant
mass cuts of the form (\ref{eq:invarmasscuts})
with $\Delta M_t \sim 15 - 35$~GeV, the matching conditions of the $(e^+e^-)(e^+e^-)$
forward scattering operators are
dominated by the NRQCD phase space contributions, i.e.\ they can be computed
from the difference 
between the (potentially) divergent NRQCD phase space integrations without any
cuts and the ones with the cuts in Eq.~(\ref{eq:invarmasscuts}) being
imposed. 
This is because using the $\overline{\mbox{MS}}$ scheme in NRQCD diagrams 
involving the unstable top propagator of Eq.~(\ref{unstableprop}) largely
overestimates the contributions from unphysical phase space 
regions that are parametrically away from the potential, soft and ultrasoft 
regions that can be described by NRQCD.
Thus the main numerical effect of the ``phase space matching'' procedure is 
obtained by removing these unphysical contributions and can be carried out within
NRQCD itself. 

An important conceptual aspect of the invariant mass cuts defined in
Eq.~(\ref{eq:invarmasscuts}) is that already for moderate cuts 
$\Delta M_{t}\sim 15-35$~GeV the cut $\Delta \bmp$ on the nonrelativistic
(anti)top three-momentum $\bmp$ is $\Delta \bmp \sim \sqrt{2m_t \Delta
  M_{t}}\sim 100$~GeV, and thus represents a hard scale of the order $m_t$. 
This justifies the implementation of the phase space effects into the matching
conditions of the Wilson coefficients. 
Since $\Lambda\equiv \sqrt{2 m_t \Delta M_t}$ is parametrically of order $m_t$ we
use for our bookkeeping the counting $\Lambda \sim m_t$.
In this counting scheme the phase space constraints are incorporated through 
the NRQCD Wilson coefficients. On the other hand, numerically   
the scales $\Delta M_t$ and $\Lambda$ are sufficiently below the
top mass scale such that all $t\bar t$ phase space configurations that pass the
invariant mass constraint can still be adequately described by NRQCD. This fact
is crucial for the phase space matching method briefly outlined in the following.

\begin{figure}[t]
  \begin{center}
  \includegraphics[width=0.9\textwidth]{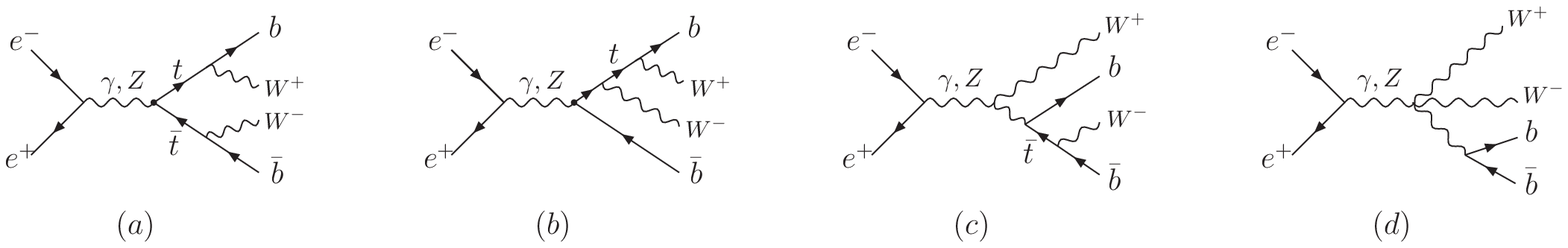}
  \caption{(a) Full theory diagram for $e^+e^-\to t\bar t \to W^+W^- b\bar b$.
   (b,c) Typical single-resonant full theory diagram for  
  $e^+e^-\to W^- t\bar b \to W^+W^- b\bar b$. 
   (d) Typical full theory diagram for $e^+e^-\to W^+W^- b\bar b$ without top or
   antitop quarks as intermediate states. 
  \label{fig:eebbWW}
}
  \end{center}
\end{figure}

Consider first the case without QCD effects ($\alpha_s=0)$. For the determination of the Wilson coefficients 
$\tilde C_{V/A}^{(n)}$ we need to know the result for the inclusive cross
section $\sigma_{\rm incl}^{\alpha_s=0}(\Lambda)$ with invariant mass
constraints~\footnote{The Wilson coefficients $\tilde C_{V/A}^{(n)}$ correspond to the $C_{4 e}^{(k)}$ coefficients
of Eq.~(\ref{eq:master}), following the notation of \cite{Hoang:2010gu}, which is explained in their Sec. II.}.
In the common approach to matching computations, 
$\sigma_{\rm incl}^{\alpha_s=0}(\Lambda)$ is computed in the full relativistic
theory. After the result is expanded nonrelativistically using the counting $v\sim \sqrt{\alpha_{\rm EW}}$,
one can identify the 
pieces belonging to the Wilson coefficients $\tilde C_{V/A}^{(n)}$. 
On the other hand, as mentioned above, the $t\bar t$ phase space regions passing
the invariant mass cuts can be determined within the nonrelativistic
expansion. We therefore write the expression for the inclusive cross section as
a sum of two terms,
\begin{align}
& \sigma_{\rm incl}^{\alpha_s=0}(\Lambda) \, = \, 
\sigma_{\rm NRQCD}^{\alpha_s=0}(\Lambda)  + 
\sigma_{\rm rem}^{\alpha_s=0}(\Lambda)
\,.
\end{align}
Here, $\sigma_{\rm NRQCD}^{\alpha_s=0}$ is the cross section computed from NRQCD
Feynman rules with the (anti)top invariant mass constraints being applied for
the phase space integration. The parameter $\Lambda$ is related to the invariant mass
cut $\Delta M_t$ and we use the formal counting $\Lambda\sim m_t$ according to
the discussion above. In this computation the $(e^+e^-)(e^+e^-)$ forward 
scattering operators do not contribute, and the resulting expressions are just
the nonrelativistic expansions of full theory squared matrix elements containing
the square of the double resonant diagram $e^+e^-\to t\bar t \to  W^+W^-b\bar b$
(see Fig.~\ref{fig:eebbWW}a)
and the interference of the double resonant diagram with the diagrams for 
$e^+e^-\to  W^+W^-b\bar b$ having only either the top or the antitop in
intermediate stages (see Fig.~\ref{fig:eebbWW}b and c for typical diagrams). 
The
contributions to the Wilson coefficients $\tilde C_{V/A}^{(n)}$ that result from 
$\sigma_{\rm NRQCD}^{\alpha_s=0}(\Lambda)$ are local (i.e. energy-independent)
and have the form 
\begin{align}
\frac{\Gamma_t}{\Lambda}\,\sum_{n,k=0} \,\left[ \left(\frac{m_t \Gamma_t}{\Lambda^2}\right)^n
\times \left( \frac{\Lambda^2}{m_t^2} \right)^k \right] 
\,,
\label{eq:PSmatching}
\end{align}
where the first term in the expansion, proportional to $\Gamma_t / \Lambda$, arises from the 
leading order diagram for the phase space matching computation (right diagram in Fig.~\ref{fig3}) and gives 
the dominant phase space correction to the $t\bar{t}$ cross section. In the
counting $\Lambda\sim m_t$, it constitutes an ${\cal O}(v^2)$ correction 
(NLL\footnote{In the NRQCD approach were logarithms are summed systematically
the counting LO, NLO$\dots$ is replaced by leading logarithmic (LL) order, next-to-leading logarithmic (NNL) order
and so on.}), since
$\Gamma_t/m_t\sim v^2$.
The
$\Lambda^2/m_t^2$ terms arise from insertions of operators that are higher
order in the nonrelativistic expansion, and since they are formally of order unity,
can lead to power-counting breaking contributions. 
However, we find that the numerical effects of the power-counting breaking
contributions are very small and do not spoil the nonrelativistic
expansion. This is partly due to the fact that the phase space cutoff $\Lambda$
is sufficiently smaller than the convergence 
radius of the nonrelativistic expansion. The phase space matching procedure
can thus be implemented for values of the invariant mass cut
in the range $m_t \Gamma_t \ll \Lambda^2 \lsim m_t^2$, but not for the total cross section 
where $\Lambda$ is numerically of order $m_t$. 

The remainder contribution of the inclusive cross section, 
$\sigma_{\rm rem}^{\alpha_s=0}(\Lambda)$ accounts for all other 
contributions to the full theory matrix element. This includes for example pure
background $e^+e^-\to W^+W^- b\bar b$ diagrams, see Fig.~\ref{fig:eebbWW}d for a
typical diagram, and also the square of the single-top diagrams in
Figs.~\ref{fig:eebbWW}b and c. One can check that the first two terms in the 
phase space matching series (\ref{eq:PSmatching}) agree with the two
first terms in the expansion in $\Lambda^2/m_t^2$ of the full NLO non-resonant result. This 
means that the remainder contributions, not calculable within the phase space matching
approach, first contribute with $\Gamma_t \Lambda^3/m_t^4$ terms, and should therefore
be small for the range of $\Lambda$ used in \cite{Hoang:2010gu}.
Indeed, the remainder contribution was determined in  \cite{Hoang:2010gu} by computing the
full  $e^+e^- \to W^+W^- b \bar{b}$ cross section at tree-level with MadGraph,
and shown to be smaller than $5$~fb; it can thus be neglected
in view of the expected experimental precision expected at a
future linear collider. 
Since no kinematic or dynamical enhancement
is expected for the QCD corrections to the remainder contributions, 
we assume these are also negligible when $\alpha_s\ne 0$. This simplifies the computations
substantially and makes the determination of higher order QCD corrections
feasible in the phase space matching approach. A calculation of the full NNLO
non-resonant contributions along the lines of Sec.~\ref{NNLOnonres} shall confirm this assumption.

The QCD corrections introduce powers of $(\alpha_s m_t/\Lambda)$ in the phase space matching
series (\ref{eq:PSmatching}).
As far as the $\alpha_s$-expansion is concerned, we have
found that the N${}^3$LL (${\cal O}(\alpha_s^2)$) corrections to the phase
space matching contributions need to be determined to meet the experimental
precision expected at a future linear collider. An important part of
these N${}^3$LL corrections have been computed analytically in~\cite{Hoang:2010gu}, together 
with the complete set of NLL and NNLL ones.
It is interesting to note that at 
N${}^3$LL order one has to include also 
the phase space matching for $(e^+e^-)(t\bar t)$ top production
operators. This is because at higher orders in the loop expansion, one has to account 
for the phase space 
matching contributions of subdiagrams to remove non-analytic matrix element
terms from the matching equations and to achieve that the matching coefficients 
are analytic in the external energy.
The 
procedure of carrying out the phase space matching is thus analogous
to the common
matching and renormalization methods for stable particle theories.

%%%%%%%%%%%%%%%
\subsection{Results}
%%%%%%%%%%%%%%%

%
\begin{figure}[t]
  \begin{center}
  \includegraphics[width=0.49\textwidth]{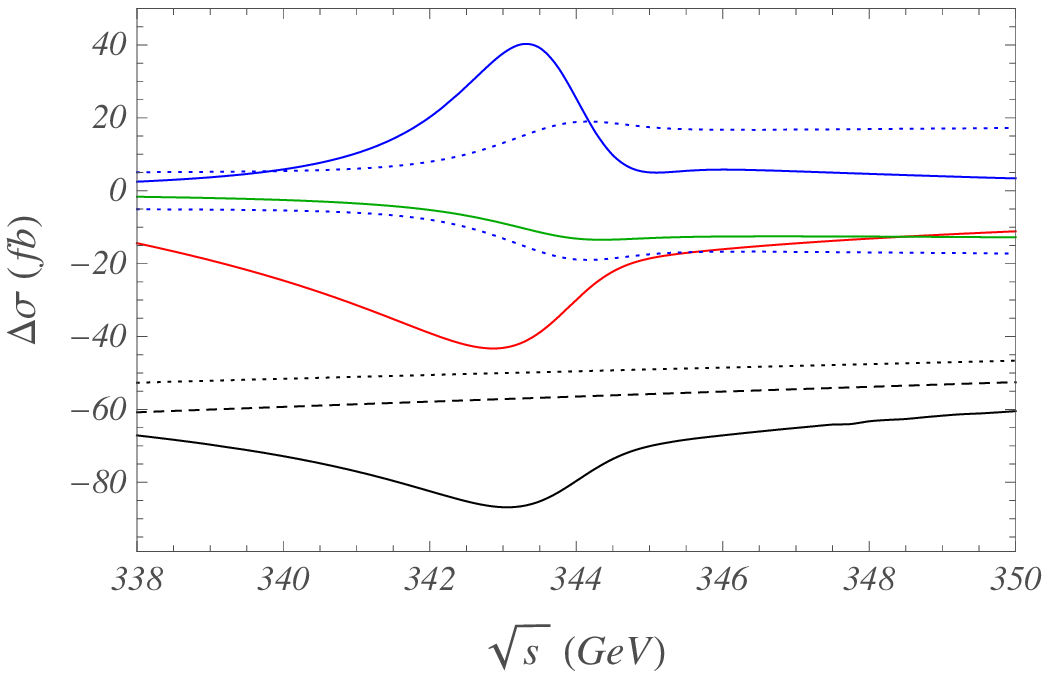}\;
  \includegraphics[width=0.49\textwidth]{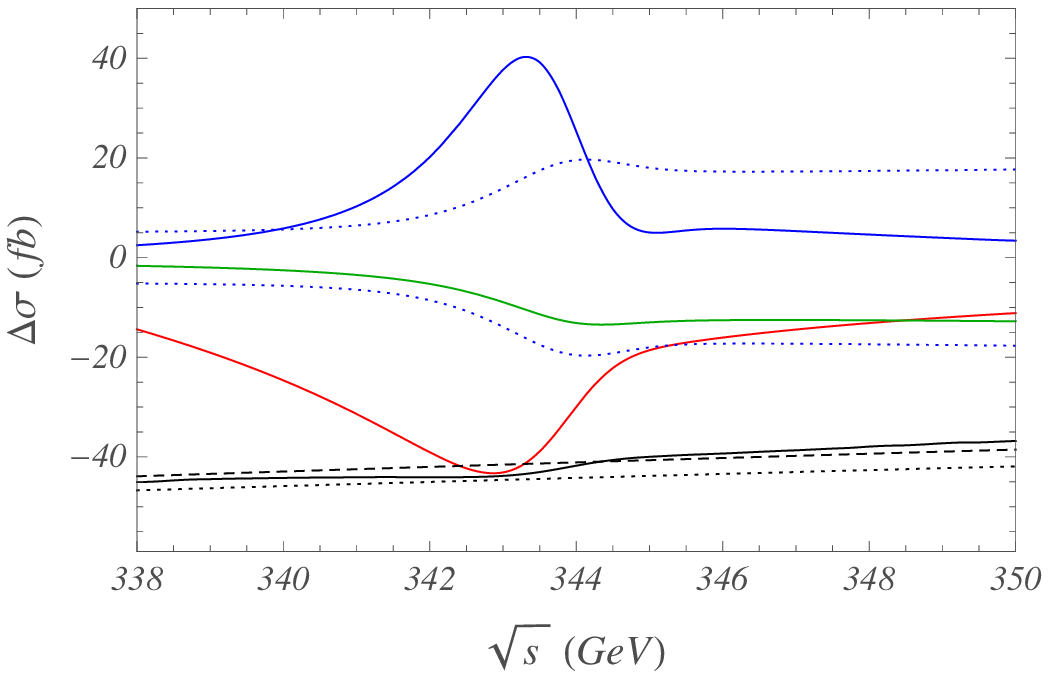}
  \caption{Sizes of the different contributions to the inclusive cross section 
arising from electroweak interactions as a function of the total c.o.m
energy for $\Delta M_t=15$~GeV (left) and $\Delta M_t=35$~GeV (right):
(green line) NNLL hard one-loop electroweak effects
from Ref.~\cite{Hoang:2006pd}, (red line) NNLL finite lifetime corrections 
from Ref.~\cite{Hoang:2004tg}, (blue line) NNLL QED effects, and phase
space matching corrections at NLL, NNLL and N${}^3$LL (dotted, dashed and
solid black lines, respectively). The blue dotted lines correspond to the expected
experimental uncertainties at the LC. The values used for
input parameters are those of Ref.~\cite{Hoang:2010gu}.
}
  \label{fig:SigmaCorrec}
  \end{center}
\end{figure}

In Fig.~\ref{fig:SigmaCorrec} we show the phase space matching corrections to
the inclusive $t\bar t$ threshold cross section up to N${}^3$LL order, for invariant
mass cuts of $\Delta M_t=15$~GeV and $\Delta M_t=35$~GeV, and compare them
with the rest of electroweak corrections. The QED
effects arise from the electromagnetic correction to the QCD Coulomb potential
and the one-loop QED matching correction to the
Wilson coefficient of the $t\bar t$ current. The hard one-loop electroweak corrections 
have been obtained in Ref.~\cite{Hoang:2006pd}.
The type-1 finite lifetime corrections represent all
finite lifetime corrections which are not related to phase space
constraints. They consist of the
corrections generated by the imaginary interference matching coefficient,
 the time dilation corrections to
the Green function, both known at NNLL order, and the contributions from
the renormalization group summation of the NLL phase space logarithms~\cite{Hoang:2004tg}.
The QED, hard electroweak and type-1 finite lifetime corrections do not depend
on phase space restrictions and are therefore identical in both panels. 
In Fig.~\ref{fig:SigmaCorrec} the blue dotted lines represent a rough estimation 
of the expected experimental uncertainties at a
future linear collider consisting of an energy-independent error of $5$~fb and a
$2$\% relative uncertainty with respect to the full prediction, both being added
quadratically.

\begin{figure}[t]
  \begin{center}
  \includegraphics[width=0.75\textwidth]{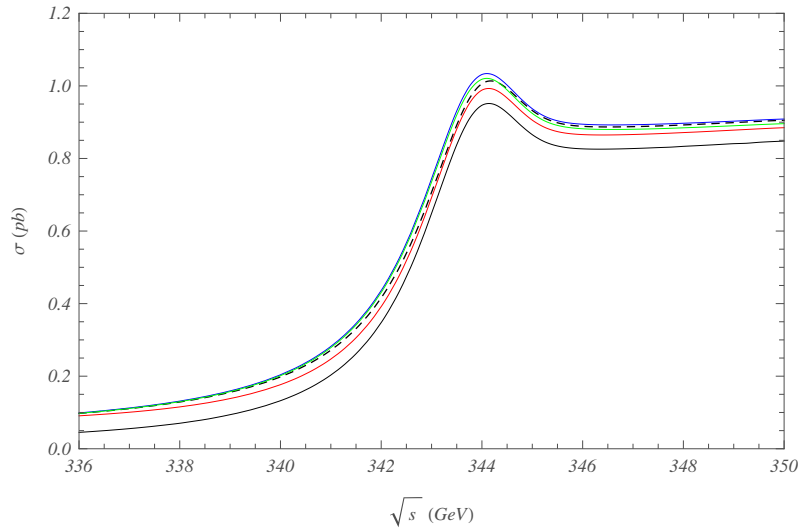}
  \caption{Total inclusive top pair production cross section from NRQCD: starting
from the pure QCD NNLL prediction (black dashed line), we add step-by-step
the QED corrections (blue line), the hard electroweak corrections (green line),
the type-1 finite lifetime corrections (red line) and the N${}^3$LL phase space
corrections (black solid line) for $\Delta M_t=35$~GeV.
}
  \label{fig:FinalQRIP35GeVCutnu02}
  \end{center}
\end{figure}

We see that the QED (blue lines) and the type-1 finite lifetime corrections (red
lines) are sizeable (at the level of $40$~fb) only in the peak region just below
$\sqrt{s}=2 m_t$. Due to their different signs
the QED corrections and the type-1 finite lifetime corrections cancel each other
to a large extent in the peak region. The hard electroweak corrections (green
lines) represent a multiplicative factor of -1.2\% to the total
cross section and are therefore very small below the peak and at the level of
$12$-$13$~fb above the peak region.
We see that the phase space matching contributions represent the largest of the
four classes of electroweak effects. In contrast to the other classes of
electroweak effects they do not decrease strongly for energies below the peak
region. For $\Delta M_t=15$~GeV the N${}^3$LL phase space matching contributions
amount between $-85$ and $-65$~fb and for $\Delta M_t=35$~GeV they are
between $-45$ and $-35$~fb. The overall size of the phase space matching
corrections decreases for larger values of the top invariant mass cut $\Delta
M_t$. We emphasize, however, that the results obtained within the phase space
matching approach are 
valid only for moderate values of $\Delta M_t$ in the region between 
$15$ and $35$~GeV. For invariant mass cuts below $15$~GeV the phase space
constraints are not related anymore to hard effects and for invariant mass cuts
substantially above $35$~GeV matching contributions that need to be computed
from full theory diagrams have to be included. 
The relatively flat behavior of the phase space matching contributions is
related to the fact that the dominant phase space matching contributions are
energy-independent. The small linear dependence on $\sqrt{s}$ is 
related to the $\sqrt{s}$ dependence of the virtual $\gamma$ and $Z$ propagators
of the basic $e^+e^-\to t\bar t$ process and the peak-like structure comes
from an imaginary phase space matching contribution to the $(e^+e^-)(t\bar t)$
top pair production operator which enters the N${}^3$LL inclusive cross section
in terms of a time-ordered product. 
The results for the NLL, NNLL
and N${}^3$LL phase space marching contributions displayed in Fig.~\ref{fig:SigmaCorrec}  show that the
expansion related to the phase space matching procedure is particularly good for
larger values of $\Delta M_t$ and still well under
control for $\Delta M_t=15$~GeV. We note that the rather small size of the NNLL
corrections (difference of black dotted and dashed lines) for $\Delta
M_t=15$~GeV arises from a 
cancellation between different independent NNLL corrections.

In Fig.~\ref{fig:FinalQRIP35GeVCutnu02} the
size of the four different types of electroweak corrections is shown for
predictions of the total inclusive cross section for $\Delta M_t=35$~GeV.
We again see that the
phase space matching contributions exceed by far the other electroweak
corrections. The phase space matching
contributions are between $-85$ and $-35$~fb for invariant mass cuts $\Delta
M_t$ between $15$ and $35$~GeV and are essential for realistic theoretical
predictions. In the peak and the continuum region ($\sqrt{s}\gsim 2m_t$) they
amount to $6$ to $10$\%. They are particularly important in the region below the
peak ($\sqrt{s}\lsim 2m_t$) where the cross section decreases and the
unphysical off-shell contributions of the NRQCD $t\bar t$ phase space become
dominant. Here the phase space matching contributions can amount to more than
$50$\%, and they ensure that the cross section has the correct physical
behavior.

\section{Acknowledgments}

I would like to thank the organizers for the excellent workshop and hospitality. I also thank B.~Jantzen for
comments on the manuscript.
This work is partly supported by the grant FPA2007-60323 and by the Spanish Consolider-Ingenio 2010
Programme CPAN (CSD2007-00042).

% ****************************************************************************
% BIBLIOGRAPHY AREA
% ****************************************************************************

\begin{footnotesize}
% IF YOU DO NOT USE BIBTEX, USE THE FOLLOWING SAMPLE SCHEME FOR THE REFERENCES
% ----------------------------------------------------------------------------

% ----------------------------------------------------------------------------

\end{footnotesize}

% ****************************************************************************
% END OF BIBLIOGRAPHY AREA
% ****************************************************************************

\end{document}